\documentclass[twocolumn,showpacs,preprintnumbers]{revtex4}%
\usepackage{amssymb}
\usepackage{amsmath}
\usepackage{graphicx}
\usepackage{dcolumn}
\usepackage{bm}
\usepackage{amsfonts}%
\providecommand{\U}[1]{\protect\rule{.1in}{.1in}}
\providecommand{\U}[1]{\protect\rule{.1in}{.1in}}
\begin{document}
\title{Devil's Staircase in an Optomechanical Cavity}
\author{Hui Wang}
\author{Yuvaraj Dhayalan}
\author{Eyal Buks}
\affiliation{Department of Electrical Engineering, Technion, Haifa 32000 Israel}
\date{\today }

\begin{abstract}
We study self-excited oscillation (SEO) in an on-fiber optomechanical cavity.
While the phase of SEO randomly diffuses in time when the laser power that is
injected into the cavity is kept constant, phase locking may occur when the
laser power is periodically modulated in time. We investigate the dependence
of phase locking on the amplitude and frequency of the laser power modulation.
We find that phase locking can be induced with a relatively low modulation
amplitude provided that the ratio between the modulation frequency and the
frequency of SEO is tuned close to a rational number of relatively low
hierarchy in the Farey tree. To account for the experimental results a one
dimensional map, which allows evaluating the time evolution of the phase of
SEO, is theoretically derived. By calculating the winding number of the one
dimensional map the regions of phase locking can be mapped in the plane of
modulation amplitude and modulation frequency. Comparison between the
theoretical predictions and the experimental findings yields a partial agreement.

\end{abstract}
\maketitle



%It is always \today, today,
%but any date may be explicitly specified

%PACS, the Physics and Astronomy
%Classification Scheme.

Optomechanical cavities \cite{Braginsky_653,
Hane_179,Gigan_67,Metzger_1002,Kippenberg_1172,Favero_104101,Marquardt2009}
are widely employed for various sensing \cite{Rugar1989, Arcizet2006a,
Forstner2012,Weig2013} and photonics
\cite{Lyshevski&Lyshevski_03,Stokes_et_al_90, Hossein_Zadeh_276,Wu_et_al_06,
MattEichenfield2007,Bahl2011,FlowersJacobs_221109} applications. Moreover,
such systems may allow experimental study of the crossover between classical
to quantum realms \cite{Thompson_72, Meystre2013,Kimble_et_al_01,
Carmon_223902, Arcizet2006, Gigan_67, Jayich_et_al_08, Schliesser_et_al_08,
Genes_et_al_08, Teufel_et_al_10,Poot_273}. The effect of radiation pressure
typically governs the optomechanical coupling (i.e. the coupling between the
electromagnetic cavity and the mechanical resonator that serves as a movable
mirror) when the finesse of the optical cavity is sufficiently high
\cite{Kippenberg_et_al_05,Rokhsari2005,
Arcizet2006,Gigan_67,Cooling_Kleckner06, Kippenberg_1172}, whereas, bolometric
effects can contribute to the optomechanical coupling when optical absorption
by the vibrating mirror is significant \cite{Metzger_1002,
Jourdan_et_al_08,Marino&Marin2011PRE, Metzger_133903, Restrepo_860,
Liberato_et_al_10,Marquardt_103901, Paternostro_et_al_06,Yuvaraj_430}.
Generally, bolometric effects are dominant in systems comprising of relatively
large mirrors, in which the thermal relaxation rate is comparable to the
mechanical resonance frequency \cite{Aubin_1018, Marquardt_103901,
Paternostro_et_al_06, Liberato_et_al_10_PRA}. These systems
\cite{Metzger_133903, Metzger_1002, Aubin_1018,
Jourdan_et_al_08,Zaitsev_046605,Zaitsev_1589} exhibit many intriguing
phenomena such as mode cooling and self-excited oscillation (SEO)
\cite{Hane_179,Kim_1454225,Aubin_1018,Carmon_223902,Marquardt_103901,Corbitt_021802,Carmon_123901,Metzger_133903,Bagheri_726}%
. It has been recently demonstrated that optomechanical cavities can be
fabricated on the tip of an optical fiber \cite{iannuzzi06, Ma2010,
iannuzzi10, iannuzzi11, Jung2011,
Butsch2012,Albri2013,Shkarin_013602,Baskin_563,Yuvaraj_210403,Shlomi_032910}.
These micron-scale devices, which can be optically actuated
\cite{Iannuzzi2013}, can be used for sensing physical parameters that affect
the mechanical properties of the suspended mirror (e.g. absorbed mass, heating
by external radiation, acceleration, etc.).%

\begin{figure}
[ptb]
\begin{center}
\includegraphics[
trim=0.585938in 0.126885in 0.718125in 0.296367in,
height=3.6382in,
width=3.3993in
]%
{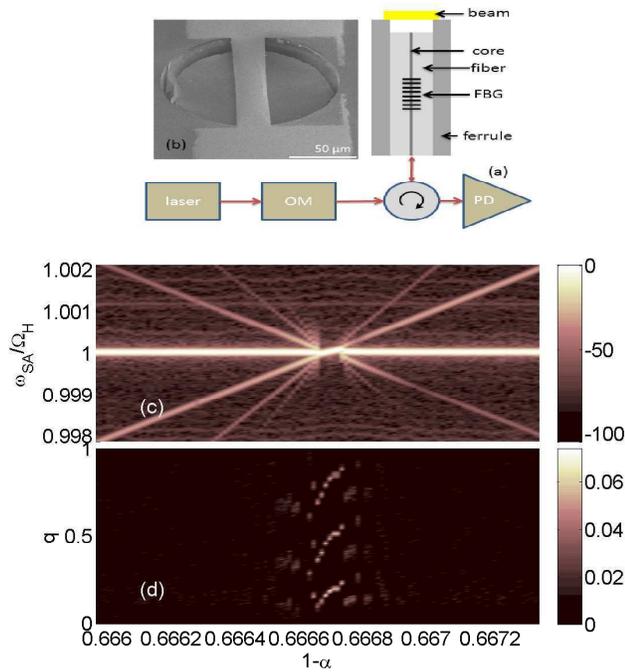}%
\caption{Experimental setup and phase locking. (a) A schematic drawing of the
sample and the experimental setup. An on-fiber optomechanical cavity is
excited by a tunable laser with external optical modulator (OM). The
mechanical resonator has quality factor $Q=\omega_{0}/2\gamma_{0}=3800$ and
the cavity has finesse $\beta_{\mathrm{F}}=2.1$. The reflected light intensity
is measured using a photodetector (PD), which is connected to both a spectrum
analyzer and an oscilloscope (not shown in the sketch). (b) Electron
micrograph of the suspended micromechanical mirror, whose mass is
$m=1.1\times10^{-12}\operatorname{kg}$. (c) Spectrum analyzer signal in dB
units vs. $1-\alpha$ and normalized measurement frequency $\omega
_{\mathrm{SA}}/\Omega_{\mathrm{H}}$. In a region near the point $1-\alpha=2/3$
entrainment occurs. The average laser power is set to $12\operatorname{mW}$,
the wavelength to $\lambda=1545.498\operatorname{nm}$ and the dimensionless
modulation amplitude to $\beta_{\mathrm{f}}=0.025$. (d) The measured
probability distribution $F\left(  q\right)  $ vs. $1-\alpha$. In the same
region where entrainment occurs [see panel(c)] the distribution$\ F\left(
q\right)  $ is peaked near three values [see panel (d)].}%
\label{Fig setup and PSD}%
\end{center}
\end{figure}
%EndExpansion

In a recent study \cite{Shlomi_032910} phase locking of SEO has been
investigated in an on-fiber optomechanical cavity, which is formed between a
fiber Bragg grating (FBG) mirror, serving as a static reflector, and a
vibrating mirror, which is fabricated on the tip of a single mode optical
fiber. In that experiment \cite{Shlomi_032910} SEO \cite{Rugar1989,
Arcizet2006a, Forstner2012,Weig2013} has been optically induced by injecting a
monochromatic laser light into the on-fiber optomechanical cavity. The
optically-induced SEO is attributed to the bolometric optomechanical coupling
between the optical mode and the mechanical resonator
\cite{Zaitsev_046605,Zaitsev_1589}. It was found in \cite{Shlomi_032910} that
the phase of the SEO can be locked by periodically modulating the laser power
that is injected into the cavity. Such phase locking results in entrainment
\cite{Hamerly_1504_04410,Blocher_835,Villanueva_177208}, i.e. synchronization
\cite{huygens1986pendulum,Rosenblum_401,Cross_224101} between the SEO and the
external modulation \cite{Georg_043603,Shah_113602}. Synchronization in
self-oscillating systems \cite{Rosenblum_401,Osipov_1,
Pikovsky_2291,Afraimovich_1,Kuznetsov_221,Landa_1,Fradkov_1}, in general, can
be the result of interaction between systems \cite{Pecora_2374, Pecora_821,
de_R7359, Fujisaka_32,Landa_414, Warminski_677,Lemonde_053602}, external noise
\cite{Pikovsky_576, Balanov_L113, Czolczynski_937, Balanov_041105, Zhang_411,
Rosenblum_1804, Pikovsky_219,Yang_1753} or other outside sources, periodic
\cite{Koronovskii_847, Nikitin_171,Min_202} or non-periodic
\cite{Nakabayashi_163,Rosenblum_264102}. Synchronization can also be activated
by applying a delayed feedback \cite{Janson_010601, Balanov_1, Scholl_281,
Hamdi_1}.

In the experiment reported in \cite{Shlomi_032910} phase locking
\cite{Anishchenko_117,Pandey_3,Paciorek_1723,Adler_351,Jensen_1637,DosSantos_1147}
has been studied for the case where the ratio between the modulation frequency
and the frequency of SEO, which is henceforth labeled as $1-\alpha$, was tuned
close to two values, the first value $1-\alpha=1$ corresponds to modulation at
the SEO frequency, and the second value $1-\alpha=2$ corresponds to modulation
at twice the SEO frequency. In the current paper we extend the study and
investigate phase locking for arbitrary values of the dimensionless parameter
$\alpha$ in the range $\left[  0,1\right]  $. This is done by experimentally
mapping the region of phase locking in the plane of the modulation amplitude
and modulation frequency. We find that phase locking can be induced with
relatively low modulation amplitude provided that $\alpha$ is tuned close to a
rational number of relatively low hierarchy in the Farey tree
\cite{Glazier_790}. To account for the experimental results we theoretically
evaluate the effect of modulation on the time evolution of the phase of SEO.
Some simplifying assumptions and approximations lead to a one dimensional map
[Eq. (\ref{q_n+1}) below], which describes the change that is accumulated over
a single period of mechanical oscillation in the relative phase between SEO
and the modulation, which is labelled by $2\pi q$. The winding number of the
one dimensional map exhibits Devil's staircase (Fig. \ref{Fig W} below)
\cite{Jensen_1637,Ben-Jacob_822,Reichhardt_414,Shim_95}, i.e. plateaus near
rational values of the parameter $\alpha$, corresponding to regions where
phase locking \cite{Paciorek_1723,Adler_351} occurs. Partial agreement is
obtained from the comparison between the experimental findings and theoretical predictions.

The optomechanical cavity, which is schematically shown in Fig.
\ref{Fig setup and PSD}(a), was fabricated on the flat polished tip of a
single mode fused silica optical fiber with outer diameter of $126%
%TCIMACRO{\unit{\U{3bc}m}}%
%BeginExpansion
\operatorname{\mu m}%
%EndExpansion
$ (Corning SMF-28 operating at wavelength band around $1550%
%TCIMACRO{\unit{nm}}%
%BeginExpansion
\operatorname{nm}%
%EndExpansion
$) held in a zirconia ferrule \cite{Yuvaraj_210403}. A $10-%
%TCIMACRO{\unit{nm}}%
%BeginExpansion
\operatorname{nm}%
%EndExpansion
$-thick chromium layer and a $200%
%TCIMACRO{\unit{nm}}%
%BeginExpansion
\operatorname{nm}%
%EndExpansion
$ gold layer were successively deposited by thermal evaporation. The bilayer
was directly patterned by a focused ion beam to the desired mirror shape ($20-%
%TCIMACRO{\unit{\U{3bc}m}}%
%BeginExpansion
\operatorname{\mu m}%
%EndExpansion
$-wide doubly clamped beam). Finally, the mirror was released by etching
approximately $12%
%TCIMACRO{\unit{\U{3bc}m}}%
%BeginExpansion
\operatorname{\mu m}%
%EndExpansion
$ of the underlying silica in 7\% HF acid ($90%
%TCIMACRO{\unit{min}}%
%BeginExpansion
\operatorname{min}%
%EndExpansion
$ etch time at room temperature). The suspended mirror remained supported by
the zirconia ferrule, which is resistant to HF.

The static mirror of the optomechanical cavity was provided by a FBG mirror
\cite{Zaitsev_046605} (made using a standard phase mask technique
\cite{Anderson_566}, with grating period of $0.527%
%TCIMACRO{\unit{\U{3bc}m}}%
%BeginExpansion
\operatorname{\mu m}%
%EndExpansion
$ and length $\approx8%
%TCIMACRO{\unit{mm}}%
%BeginExpansion
\operatorname{mm}%
%EndExpansion
$) having reflectivity band of $0.4%
%TCIMACRO{\unit{nm}}%
%BeginExpansion
\operatorname{nm}%
%EndExpansion
$ full width at half maximum (FWHM) centered at $1545%
%TCIMACRO{\unit{nm}}%
%BeginExpansion
\operatorname{nm}%
%EndExpansion
$. The length of the optical cavity was $l\approx10%
%TCIMACRO{\unit{mm}}%
%BeginExpansion
\operatorname{mm}%
%EndExpansion
$, providing a free spectral range of $\Delta\lambda=\lambda_{0}%
^{2}/2n_{\mathrm{eff}}l\approx80$ pm (where $n_{\mathrm{eff}}$ $=1.468$ is the
effective refraction index for SMF-28). Monochromatic light was injected into
the fiber bearing the cavity on its tip from a laser source with an adjustable
output wavelength $\lambda$ and power level $P_{\mathrm{L}}$. The laser was
connected through an optical circulator, that allowed the measurement of the
reflected light intensity $P_{\mathrm{R}}$ by a fast responding photodetector
[see Fig. \ref{Fig setup and PSD}(a)]. The detected signal was analyzed by an
oscilloscope and a spectrum analyzer. The experiments were performed in vacuum
(at residual pressure below $0.01%
%TCIMACRO{\unit{Pa}}%
%BeginExpansion
\operatorname{Pa}%
%EndExpansion
$) at a base temperature of $77%
%TCIMACRO{\unit{K}}%
%BeginExpansion
\operatorname{K}%
%EndExpansion
$. The laser power and laser wavelength were first tuned into the regime of
SEO before the modulation was turned on.

Phase locking has been measured near all fractions $\alpha=n_{1}/n_{2}$ in the
range $0<\alpha<1$, where $n_{2}\in\left\{  1,2,3,4,5\right\}  $. The case
$1-\alpha=2/3$ is demonstrated by Fig. \ref{Fig setup and PSD}. The plot in
panel (c) exhibits the measured signal of a spectrum analyzer, which is
connected to the photodetector, vs. normalized modulation frequency $1-\alpha$
and normalized measurement frequency $\omega_{\mathrm{SA}}/\Omega_{\mathrm{H}%
}$, where $\Omega_{\mathrm{H}}/2\pi=236.3%
%TCIMACRO{\unit{kHz}}%
%BeginExpansion
\operatorname{kHz}%
%EndExpansion
$ is the frequency of SEO. In the region of phase locking near the point
$1-\alpha=2/3$ the spectral peak corresponding to SEO coincides with the
sideband corresponding to the power modulation. The other spectral lines in
panel (c) converging to the central point $1-\alpha=2/3$ and $\omega
_{\mathrm{SA}}/\Omega_{\mathrm{H}}=1$ represent higher order products of
frequency mixing between $\Omega_{\mathrm{H}}$ and $\left(  1-\alpha\right)
\Omega_{\mathrm{H}}$ \cite{Antoni_68005}. The plot in panel (d) of Fig.
\ref{Fig setup and PSD} exhibits the measured probability distribution
$F\left(  q\right)  $ of the variable $q$, which represents the relative phase
between SEO and the modulation in units of $2\pi$. The distribution $F\left(
q\right)  $ is extracted from the oscilloscope's data by employing the
zero-crossing technique \cite{Cutler_136}. While $F\left(  q\right)  $ is
found to have a nearly uniform distribution away from the point $1-\alpha
=2/3$, three pronounced peaks are observed in the region of phase locking near
that point, suggesting that the relative phase undergoes a limit cycle of
period $3$.

To account for the experimental findings we theoretically investigate under
what conditions phase locking of SEO is expected to occur. In the limit of
small displacement the dynamics of the system can be approximately described
using a single evolution equation \cite{Zaitsev_1589,Dykman_1646}. The
theoretical model that is used to derive the evolution equation [see Eq.
(\ref{A dot}) below] is briefly described below. Note that some optomechanical
effects that were taken into account in the theoretical modeling
\cite{Zaitsev_1589} were found experimentally to have a negligible effect on
the dynamics \cite{Zaitsev_046605} (e.g. the effect of radiation pressure). In
what follows such effects are disregarded.

The micromechanical mirror in the optical cavity is treated as a mechanical
resonator with a single degree of freedom $x$ having mass $m$ and linear
damping rate $\gamma_{0}$ (when it is decoupled from the optical cavity). It
is assumed that the angular resonance frequency of the mechanical resonator
depends on the temperature $T$ of the suspended mirror. For small deviation of
$T$ from the base temperature $T_{0}$ (i.e. the temperature of the supporting
substrate) it is taken to be given by $\omega_{0}-\beta T_{\mathrm{R}}$, where
$T_{\mathrm{R}}=T-T_{0}$ and where $\beta$ is a constant. Furthermore, to
model the effect of thermal deformation \cite{Metzger_133903} it is assumed
that a temperature dependent force given by $m\theta T_{\mathrm{R}}$, where
$\theta$ is a constant, acts on the mechanical resonator \cite{Yuvaraj_430}.
When noise is disregarded, the equation of motion governing the dynamics of
the mechanical resonator is taken to be given by%
\begin{equation}
\frac{\mathrm{d}^{2}x}{\mathrm{d}t^{2}}+2\gamma_{0}\frac{\mathrm{d}%
x}{\mathrm{d}t}+\left(  \omega_{0}-\beta T_{\mathrm{R}}\right)  ^{2}x=\theta
T_{\mathrm{R}}\;. \label{x eom}%
\end{equation}

The intra-cavity optical power incident on the suspended mirror is denoted by
$P_{\mathrm{L}}I\left(  x\right)  $, where $P_{\mathrm{L}}$ is the injected
laser power, and the function $I\left(  x\right)  $ depends on the mechanical
displacement $x$ [see Eq. (\ref{I(x)}) below]. The time evolution of the
relative temperature $T_{\mathrm{R}}$ is governed by the thermal balance
equation%
\begin{equation}
\frac{\mathrm{d}T_{\mathrm{R}}}{\mathrm{d}t}=Q-\kappa T_{\mathrm{R}}\;,
\label{T_R eom}%
\end{equation}
where $Q=\eta P_{\mathrm{L}}I\left(  x\right)  $ is proportional to the
heating power, $\eta$ is the heating coefficient due to optical absorption and
$\kappa$ is the thermal decay rate.

The function $I\left(  x\right)  $ depends on the properties of the optical
cavity that is formed between the suspended mechanical mirror and the on-fiber
static reflector. The finesse of the optical cavity is limited by loss
mechanisms that give rise to optical energy leaking out of the cavity. The
main escape routes are through the on-fiber static reflector, through
absorption by the metallic mirror, and through radiation. The corresponding
transmission probabilities are respectively denoted by $\mathcal{T}%
_{\mathrm{B}}$, $\mathcal{T}_{\mathrm{A}}$ and $\mathcal{T}_{\mathrm{R}}$. In
terms of these parameters, the function $I\left(  x\right)  $ is given by
\cite{Zaitsev_046605}%
\begin{equation}
I\left(  x\right)  =\frac{\beta_{\mathrm{F}}\left(  1-\frac{\beta_{-}^{2}%
}{\beta_{+}^{2}}\right)  \beta_{+}^{2}}{1-\cos\frac{4\pi x_{\mathrm{D}}%
}{\lambda}+\beta_{+}^{2}}\;, \label{I(x)}%
\end{equation}
where $x_{\mathrm{D}}=x-x_{\mathrm{R}}$ is the displacement of the mirror
relative to a point $x_{\mathrm{R}}$, at which the energy stored in the
optical cavity in steady state obtains a local maximum, $\beta_{\pm}%
^{2}=\left(  \mathcal{T}_{\mathrm{B}}\pm\mathcal{T}_{\mathrm{A}}\pm
\mathcal{T}_{\mathrm{R}}\right)  ^{2}/8$ and where $\beta_{\mathrm{F}}$ is the
cavity finesse. The reflection probability $R_{\mathrm{C}}=P_{\mathrm{R}%
}/P_{\mathrm{L}}$ is given in steady state by \cite{Yurke_5054,Zaitsev_046605}
$R_{\mathrm{C}}=1-I\left(  x\right)  /\beta_{\mathrm{F}}$. For sufficiently
small $x$, the expansion $I\left(  x\right)  =I_{0}+I_{0}^{\prime}x+\left(
1/2\right)  I_{0}^{\prime\prime}x^{2}+O\left(  x^{3}\right)  $ can be
employed, where a prime denotes differentiation with respect to the
displacement $x$.%

\begin{figure}
[ptb]
\begin{center}
\includegraphics[
height=2.8449in,
width=3.4537in
]%
{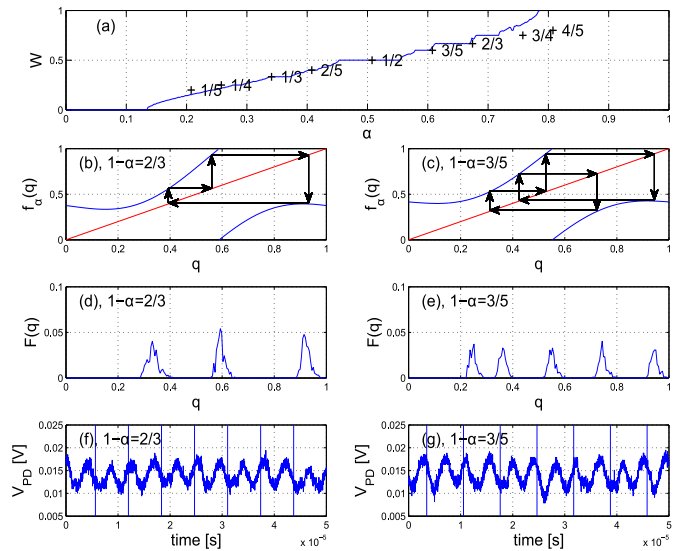}%
\caption{The winding number and limit cycles. (a) Devil's staircase in the
plot of the winding number $W$ vs. $\alpha$ for the case where $\beta
_{\mathrm{f}}=0.0355$. Locally stable limit cycles for the case $1-\alpha=2/3$
and $\beta_{\mathrm{f}}=0.025$ are presented in panels (b), (d) and (f) and
for the case $1-\alpha=3/5$ and $\beta_{\mathrm{f}}=0.028$ in panels (c), (e)
and (g). The map $f_{\alpha}\left(  q\right)  $ together with the
corresponding limit cycle for the case $1-\alpha=2/3$ ($1-\alpha=3/5$) is
depicted in panel (b) [panel (c)], the experimentally measured probability
distribution $F\left(  q\right)  $ in panel (d) [panel (e)], and a sample
temporal data in panel (f) [panel (g)] (the vertical lines label the beginning
points of each modulation period). In the experimental measurements presented
in this plot a pulse power modulation having a rectangular (instead of a
sinusoidal) waveform has been employed. For that case the recursive relation
$q_{n+1}=f_{\alpha}\left(  q_{n}\right)  $ is derived by first performing a
Fourier decomposition of the rectangular waveform, and then calculating the
contribution of each Fourier component using Eq. (\ref{f(q_n)=}).}%
\label{Fig W}%
\end{center}
\end{figure}
%EndExpansion

Consider the case where the laser power $P_{\mathrm{L}}$ is modulated in time
according to $P_{\mathrm{L}}\left(  t\right)  =P_{0}+P_{1}\left(  t\right)  $,
where $P_{0}$ is a constant and $P_{1}\left(  t\right)  $ is assumed to have a
vanishing average. When both $P_{1}$ and $I-I_{0}$ are sufficiently small, the
problem can be significantly simplified by employing the approximation
$Q\simeq\eta P_{0}I+\eta P_{1}I_{0}$. The displacement $x\left(  t\right)  $
is expressed in terms of the complex amplitude $A$ as $x\left(  t\right)
=x_{0}+2\operatorname{Re}A$, where $x_{0}$, which is given by $x_{0}%
=\eta\theta P_{0}I_{0}/\kappa\omega_{0}^{2}$, is the averaged
optically-induced static displacement. For a small displacement, the evolution
equation for the complex amplitude $A$ is found to be given by
\cite{Zaitsev_1589,Shlomi_032910}%
\begin{equation}
\dot{A}+\left(  \Gamma_{\mathrm{eff}}+i\Omega_{\mathrm{eff}}\right)
A=\xi\left(  t\right)  +\vartheta\left(  t\right)  \;, \label{A dot}%
\end{equation}
where both the effective resonance frequency $\Omega_{\mathrm{eff}}$ and the
effective damping rate $\Gamma_{\mathrm{eff}}$ are real even functions of
$\left\vert A\right\vert $. To second order in $\left\vert A\right\vert $ they
are given by $\Gamma_{\mathrm{eff}}=\Gamma_{0}+\Gamma_{2}\left\vert
A\right\vert ^{2}$ and $\Omega_{\mathrm{eff}}=\Omega_{0}+\Omega_{2}\left\vert
A\right\vert ^{2}$, where $\Gamma_{0}=\gamma_{0}+\eta\theta P_{1}I_{0}%
^{\prime}/2\omega_{0}^{2}$, $\Gamma_{2}=\gamma_{2}+\eta\beta P_{1}%
I_{0}^{\prime\prime}/4\omega_{0}$, $\gamma_{2}$ is the intrinsic mechanical
nonlinear quadratic damping rate \cite{Zaitsev_859}, $\Omega_{0}=\omega
_{0}-\eta\beta P_{1}I_{0}/\kappa$ and $\Omega_{2}=-\eta\beta P_{1}%
I_{0}^{\prime\prime}/\kappa$. Note that the above expressions for
$\Gamma_{\mathrm{eff}}$ and $\Omega_{\mathrm{eff}}$ are obtained by making the
following assumptions: $\kappa^{2}/\omega_{0}^{3}\lambda\ll\beta/\theta
\ll1/2\omega_{0}x_{0}$ and $\kappa\ll\omega_{0}$, both typically hold
experimentally \cite{Zaitsev_046605}. The term $\xi\left(  t\right)  $, which
is given by $\xi\left(  t\right)  =\Omega_{0}^{-1}\theta T_{\mathrm{R}%
1}\left(  t\right)  $, where $T_{\mathrm{R}1}\left(  t\right)  $ is found by
solving Eq. (\ref{T_R eom}) for the case where the the laser power is taken to
be $P_{1}\left(  t\right)  $, represents the thermal force that is generated
due to the power modulation. The fluctuating term
\cite{Risken_Fokker-Planck,Fong_023825} $\vartheta\left(  t\right)
=\vartheta_{x}\left(  t\right)  +i\vartheta_{y}\left(  t\right)  $, where both
$\vartheta_{x}$ and $\vartheta_{y}$ are real, represents white noise and the
following is assumed to hold: $\left\langle \vartheta_{x}\left(  t\right)
\vartheta_{x}\left(  t^{\prime}\right)  \right\rangle =\left\langle
\vartheta_{y}\left(  t\right)  \vartheta_{y}\left(  t^{\prime}\right)
\right\rangle =2\Theta\delta\left(  t-t^{\prime}\right)  $ and $\left\langle
\vartheta_{x}\left(  t\right)  \vartheta_{y}\left(  t^{\prime}\right)
\right\rangle =0$, where $\Theta=\gamma_{0}k_{\mathrm{B}}T_{\mathrm{eff}%
}/4m\omega_{0}^{2}$, $k_{\mathrm{B}}$ is the Boltzmann's constant and
$T_{\mathrm{eff}}$ is the effective noise temperature.

In the absence of laser modulation, i.e. when $P_{1}=0$, the equation of
motion (\ref{A dot}) describes a van der Pol oscillator \cite{Pandey_3}.
Consider the case where $\Gamma_{2}>0$, for which a supercritical Hopf
bifurcation occurs when the linear damping coefficient $\Gamma_{0}$ vanishes.
Above threshold, i.e. when $\Gamma_{0}$ becomes negative, the amplitude
$A_{r}=\left\vert A\right\vert $ of SEO is given by $A_{r0}=\sqrt{-\Gamma
_{0}/\Gamma_{2}}$ and the angular frequency $\Omega_{\mathrm{H}}$ of SEO by
$\Omega_{\mathrm{H}}=\Omega_{\mathrm{eff}}\left(  A_{r0}\right)  $. For our
experimental parameters $\left\vert \Omega_{2}\right\vert \ll\omega_{0}%
/A_{r0}^{2}$, and consequently to a good approximation the dependence of
$\Omega_{\mathrm{H}}$ on $A_{r0}$ can be disregarded.%

%TCIMACRO{\FRAME{ftbpFU}{3.4546in}{2.8209in}{0pt}{\Qcb{Arnold tongue near the
%point $\alpha=1/2$. The colormap represents the measured value of the
%derivative $\mathrm{d}W/\mathrm{d}\alpha$ vs. $\alpha$ and $\beta_{\mathrm{f}%
%}$. The black dotted line represents the theoretically calculated bifurcation
%line $\beta_{\mathrm{f}}=\left(  81/128\pi\right)  ^{1/3}\left(
%\alpha-1/2\right)  ^{2/3}$. Laser parameters are the same as those listed in
%the caption of Fig. \ref{Fig setup and PSD}. The experimental value of the
%dimensionless parameter $\beta_{\mathrm{f}}$ is determined using the following
%device parameter $\Omega_{\mathrm{H}}^{3}A_{r0}/\theta\eta I_{0}=0.10\unit{W}%
%$.}}{\Qlb{Fig Arnold Tongue}}{fig3.eps}{\special{ language "Scientific Word";
%type "GRAPHIC";  maintain-aspect-ratio TRUE;  display "ICON";
%valid_file "F";  width 3.4546in;  height 2.8209in;  depth 0pt;
%original-width 10.1481in;  original-height 8.2745in;  cropleft "0";
%filename 'Fig3.eps';file-properties "XNPEU";}} }%
%BeginExpansion
\begin{figure}
[ptb]
\begin{center}
\includegraphics[
height=2.8209in,
width=3.4546in
]%
{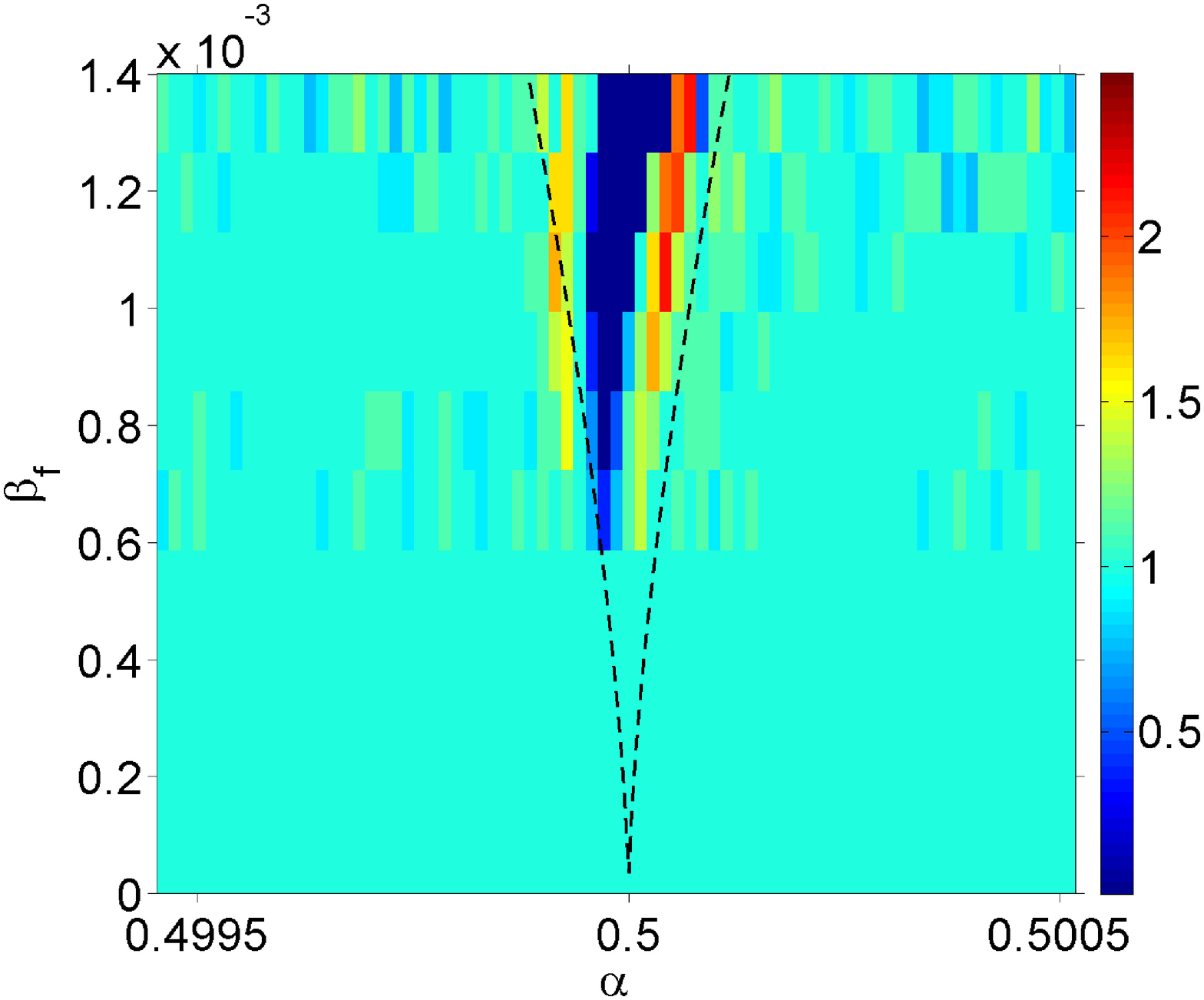}%
\caption{Arnold tongue near the point $\alpha=1/2$. The colormap represents
the measured value of the derivative $\mathrm{d}W/\mathrm{d}\alpha$ vs.
$\alpha$ and $\beta_{\mathrm{f}}$. The black dotted line represents the
theoretically calculated bifurcation line $\beta_{\mathrm{f}}=\left(
81/128\pi\right)  ^{1/3}\left(  \alpha-1/2\right)  ^{2/3}$. Laser parameters
are the same as those listed in the caption of Fig. \ref{Fig setup and PSD}.
The experimental value of the dimensionless parameter $\beta_{\mathrm{f}}$ is
determined using the following device parameter $\Omega_{\mathrm{H}}^{3}%
A_{r0}/\theta\eta I_{0}=0.10\operatorname{W}$.}%
\label{Fig Arnold Tongue}%
\end{center}
\end{figure}
%EndExpansion

The laser power modulation $P_{1}\left(  t\right)  $ is taken to be time
periodic with angular frequency $\left(  1-\alpha\right)  \Omega_{\mathrm{H}}%
$, a sinusoidal waveform and an amplitude $P_{\mathrm{p}}$, which is expressed
as $P_{\mathrm{p}}=\beta_{\mathrm{f}}\Omega_{\mathrm{H}}^{3}A_{r0}/\theta\eta
I_{0}$, where both $\alpha$ and $\beta_{\mathrm{f}}$ are real dimensionless
constants. Let $2\pi q_{n}$ be the relative phase of SEO with respect to the
external modulation after integer number $n$ of periods of mechanical
oscillation. Integrating Eq. (\ref{A dot}) over a single period of SEO (and
disregarding the noise term) yields for the case where $\beta_{\mathrm{f}}%
\ll1$ a recursive relation between $q_{n+1}$ and $q_{n}$, which reads%
\begin{equation}
q_{n+1}=f_{\alpha}\left(  q_{n}\right)  \;, \label{q_n+1}%
\end{equation}
where%
\begin{equation}
f_{\alpha}\left(  q\right)  =q+\alpha+\frac{2\beta_{\mathrm{f}}\sin\left(
\pi\alpha\right)  \cos\left(  \pi\alpha+2\pi q\right)  }{\left(
1-\alpha\right)  \left[  1-\left(  1-\alpha\right)  ^{2}\right]  }\;.
\label{f(q_n)=}%
\end{equation}

The winding number $W$ is defined by \cite{Jensen_1637}%
\begin{equation}
W=\lim_{n\rightarrow\infty}\frac{q_{n+1}-q_{1}}{n}\;. \label{W def}%
\end{equation}
For the case of a limit cycle, the winding number is a rational number given
by $W=n_{1}/n_{2}$, where $n_{2}$ is the period of the cycle and $n_{1}$ is
the number of sweeps through the unit interval $[0,1]$ in a cycle when the
mapping (\ref{f(q_n)=}) is considered as modulo $1$. Devil's staircase can be
seen in Fig. \ref{Fig W}(a), in which the winding number $W$ is plotted as a
function of $\alpha$ for the case where $\beta_{\mathrm{f}}=0.0355$. Locally
stable limit cycles are presented in Fig. \ref{Fig W} for the case
$1-\alpha=2/3$ and $\beta_{\mathrm{f}}=0.025$ [panels (b), (d) and (f)] and
for the case $1-\alpha=3/5$ and $\beta_{\mathrm{f}}=0.028$ [panels (c), (e)
and (g)]. The map $f_{\alpha}\left(  q\right)  $ together with the
corresponding limit cycle for the case $1-\alpha=2/3$ ($1-\alpha=3/5$) is
depicted in panel (b) [panel (c)], the experimentally measured probability
distribution $F\left(  q\right)  $ in panel (d) [panel (e)], and a sample
temporal data in panel (f) [panel (g)]. The comparison between the values of
$q$ corresponding to the peaks in the measured distribution $F\left(
q\right)  $ [panels (d) and (e)] and the values of $q$ corresponding to the
calculated limit cycle [panels (b) and (c)] exhibits a partial agreement.

Regions of phase locking in the plane that is spanned by the modulation
parameters (frequency and amplitude) are commonly called Arnold tongues. The
Arnold tongue near $\alpha=1/2$ is seen in Fig. \ref{Fig Arnold Tongue}. The
colormap exhibits the measured value of the derivative $\mathrm{d}%
W/\mathrm{d}\alpha$ vs. $\alpha$ and $\beta_{\mathrm{f}}$. Near the point
$\alpha=1/2$ the Arnold tongue (the region where $\mathrm{d}W/\mathrm{d}%
\alpha\simeq0$ in Fig. \ref{Fig Arnold Tongue}) represents the stability zone
of a fixed point of the second order map $\mathcal{F}_{2,\alpha}\left(
q\right)  \equiv f_{\alpha}\left(  f_{\alpha}\left(  q\right)  \right)  $. To
account for the experimental results the behavior of the map $\mathcal{F}%
_{2,\alpha}\left(  q\right)  $ is theoretically investigated near the point
$\alpha=1/2$. By expressing $\alpha$ as $\alpha=1/2+\epsilon$ and by expanding
$\mathcal{F}_{2,\alpha}\left(  q\right)  $ up to first order in $\epsilon$ the
fixed point, which is labeled as $q_{\mathrm{f}}$, can be analytically
evaluated to lowest nonvanishing order in $\beta_{\mathrm{f}}$. The region of
stability, which is found from the requirement that $\left\vert \mathcal{F}%
_{2,\alpha}^{\prime}\left(  q_{\mathrm{f}}\right)  \right\vert <1$, yields the
bifurcation line in the plane spanned by $\alpha$ and $\beta_{\mathrm{f}}$,
which is found to be given by $\beta_{\mathrm{f}}=\left(  81/128\pi\right)
^{1/3}\epsilon^{2/3}$ (see the black dotted line in Fig.
\ref{Fig Arnold Tongue}). The comparison between data and theory yields a
moderate agreement.

In summary, Devil's staircase in an on-fiber optomechanical cavity is
investigated. The device under study can be employed as a sensor operating in
the region of SEO. Future study will address the possibility of reducing phase
noise by inducing phase locking in order to enhance sensor's performance.

This work was supported by the Israel Science Foundation, the bi-national
science foundation, the Security Research Foundation in the Technion, the
Israel Ministry of Science and the Russell Berrie Nanotechnology Institute.

%Just because of unusual number of tables stacked at end
\bibliographystyle{ieeepes}
\bibliography{acompat,Eyal_Bib}
%Produces the bibliography via BibTeX.

\end{document}